\documentclass[preprint,aps,showpacs]{revtex4}
\usepackage{amssymb}
\usepackage{amsmath}
\usepackage{bm,dcolumn}

\input{tcilatex}

\begin{document}

\title{Structure and Orientation of the Moving Vortex Lattice in Clean Type
II Superconductors}
\author{$^{1,3}$Dingping Li, $^{2,3}$Andrey M. Malkin and $^{3,4}$Baruch
Rosenstein}
\affiliation{$^{1}$\textit{Department of Physics, Peking University, Beijing 100871, China%
}}
\affiliation{$^{2}$\textit{Institute of Applied Physics, Russian Academy of Science,
Nizhnii Novgorod 603600, Russia}}
\affiliation{$^{3}$\textit{National Center for Theoretical Sciences} \textit{and} \textit{%
Electrophysics Department, National Chiao Tung University,Hsinchu 30050,}
\textit{Taiwan, R. O. C}.}
\affiliation{$^{4}$\textit{Department of Condensed Matter Physics, Weizmann Institute of
Science, Rehovot 76100, Israel}}
\date{\today}

\begin{abstract}
The dynamics of moving vortex lattice is considered in the framework of the
time dependent Ginzburg - Landau equation neglecting effects of pinning. At
high flux velocities the pinning dominated dynamics is expected to cross
over into the interactions dominated dynamics for very clean materials
recently studied experimentally. The stationary lattice structure and
orientation depend on the flux flow velocity. For relatively velocities $%
V<V_{c}=\sqrt{8\pi B/\Phi _{0}}/\gamma ,$ where $\gamma $ is
inverse diffusion constant in time dependent Ginzburg - Landau
equation, vortex lattice has \ a different orientation than for
$V>V_{c}$. The two orientations can be described as motion "in
channels" and motion of "lines of vortices perpendicular to the
direction of motion. Although we start from the lowest Landau
level approximation, corrections to conductivity and the vortex
lattice energy dissipation from higher Landau levels are
systematically calculated and compared to a recent experiment.
\end{abstract}

\maketitle

\section{Introduction}

The static Abrikosov flux lattice has been experimentally observed since
sixties by great variety of techniques and lateral correlations have been
clearly observed recently up to tens of thousands of lattice spacings \cite%
{Lieber}. The remarkable advances in decoration, small-angle neutron
scattering and muon spin rotation techniques allowed recently direct glimpse
into the structure of the moving Abrikosov vortex systems \cite%
{Yaron,Kes,Pardo,Forgan}. It shows that at small flux flow velocities\
vortices move in channels as predicted in \cite{Balents}. When the flux flow
velocity increases beyond the one corresponding to the critical current, one
observes a relatively well correlated hexagonal lattice. The channels and
the plastic flow at relatively low velocities are explained by influence of
pinning on the basis of theoretical arguments \cite{Koshelev} and confirmed
by numerous simulations \cite{Koshelev,Zimanyi1,Reichhardt,Fangohr,Zimanyi2}%
. At high velocity of the moving lattice (corresponding to high electric
field), the influence of disorder is expected to diminish and a
\textquotedblleft moving Bragg glass\textquotedblright\ appears \cite%
{Giamarchi,Koshelev}. Indeed Bragg peaks roughly at positions of the
hexagonal lattice were observed \cite{Forgan} recently.

Since the theoretical prediction of the moving Bragg glass exhibiting
transverse peak effect \cite{Giamarchi}, a lot of effort has been put into
the simulation of the high driving force phase of the moving vortex system
\cite{Reichhardt,Fangohr,Zimanyi2}. In particular it was found \cite%
{Reichhardt} that as the driving force increases (or disorder decreases) the
vortex lattice suddenly changes orientation for a period of time and then
returns to a \textquotedblleft regular\textquotedblright\ drift mode. The
main emphasis in these studies mentioned above is still the effects of
pinning on the moving lattice.

Experiments at low (below $100G$) magnetic field and slow flux moving
velocity (of order $\mu m/sec$) showed that the orientation of the moving
vortex lattice is tied to the direction of motion, namely, when nearly
hexagonal lattice is observed, one always observes the orientation depicted
on Fig. 1a, never the \textquotedblleft rotated\textquotedblright\ one of
the Fig.1b \cite{Pardo}. Here the effect of pinning cannot be ignored and
plays an important role in the orientation of the vortex lattice. However
the most recent small-angle neutron scattering and muon spin rotation
experiment can probe the moving lattice at much higher velocities of order $%
cm/sec$ or even higher. The results about the orientation of the moving
lattice obtained in \cite{Forgan} seem to be different from the case at low
magnetic field and slow flux moving velocity.

The effect of pinning is expected to be smaller at higher velocities.
Alternatively one can ask what happens in very clean materials. A recent
experiment in $Pn-In$ seems to belong to this category \cite{Forgan}. As the
pinning influence diminishes with increasing flux velocity, it is natural to
ask what happen in the limit of the highest possible flux velocity (of
course, eventually the electric field destroys superconductivity, so that
the mathematical limit of the infinite driving force is unphysical)
disregarding pinning altogether.

The question of the orientation of the vortex lattice usually does not arise
in the static case. Without external electric field singling out a
particular direction one has a complete degeneracy of possible orientations
of the hexagonal vortex lattice. This is not surprising for a sufficiently
symmetric material (like $NbSe_{2}$ frequently used in experiments belongs
to this category): the rotational symmetry ensures that the free energy is
independent of the hexagonal lattice orientation. The rotational symmetry is
broken by the motion of fluxons as was confirmed experimentally \cite%
{Pardo,Andrei}. Naturally one could ask whether the particular lattice
orientation observed for example in \cite{Pardo} is necessarily tied to
pinning or might appear in clean superconductors as well. Furthermore, the
lattice also can be deformed though the deformation apparently is very small
(see Fig.1c,d in \cite{Pardo}). Is there a deformation even before pinning
centers disorder the lattice?

It would be difficult to address the question of the moving vortex lattice
structure using phenomenological models like the elastic medium \cite%
{Giamarchi} (in which individual vortices are simply not \textquotedblleft
seen\textquotedblright )\ or approximating vortices in the London
approximation by interacting lines or points $r_{i}$ in 2D \cite{Zimanyi2}.
To give an example of the problems in the London limit, let us consider
equations of motion for vortices. The driving force $\mathbf{F}$ is the
Lorentz force and the dynamics is assumed overdamped:
\begin{equation}
\eta \frac{d\mathbf{r}_{i}}{dt}=-\sum_{j\neq i}\mathbf{\nabla }U\left(
\mathbf{r}_{i}-\mathbf{r}_{j}\right) +\mathbf{F,}
\end{equation}%
where $U\left( \mathbf{r}_{i}-\mathbf{r}_{j}\right) $ is the inter - vortex
repulsive potential. The solution of these equations in the absence of
pinning is obvious: \ the \textquotedblleft boosted\textquotedblright\
hexagonal lattice of any orientation irrespective of the direction of $%
\mathbf{F}$. Thus the orientation of the lattice depends solely on initial
conditions, at least in the clean case. Therefore the approximations made in
the above phenomenological approaches are too strong.

In this paper we use the time dependent Ginzburg - Landau (TDGL) model to
study the vortex motion and structure. The TDGL approach has been remarkably
successful in describing various thermodynamical and transport properties
\cite{Brandt}. Progress in obtaining the theoretical results from the model
can be achieved only when certain additional assumptions are made. One of
the often made additional assumption is that only the lowest Landau level
(LLL) significantly contributes to physical quantities of interest. The LLL
approximation is valid for $H>\frac{H_{c2}(T)}{13}$ in the static limit \cite%
{LiHLL}. Although most of the experiments concerning moving lattice were
performed at field far below the static $H_{c2}(T)$, it has been shown long
time ago \cite{Maki,Hu} that in the presence of electric field $E$ the
effective $H_{c2}(T,E)=H_{c2}(T)-\gamma ^{2}V^{2}\Phi _{0}/(8\pi )$ where $%
V=cE/B$ is the velocity of fluxons and $\gamma $ is the inverse diffusion
constant setting the time scale in TDGL approach. This field $H_{c2}(T,E)$
could be much smaller at not very small fluxon velocities (electric field
suppresses superconductivity even more effectively than the magnetic field).
Therefore effectively one can move into the region of validity of the LLL
approximation at sufficiently large currents. Moreover one expects that,
even beyond the region of validity of the LLL approximation, physics is
qualitatively the same.

We solve TDGL equations for a moving vortex solid without disorder and find
the vortex structure to which the moving lattice relaxes irrespective of
initial conditions \cite{Maki,Hu,Troy}. It turns out that the preferred
lattice is rhombic. The distortion is velocity dependent. Remarkably the
orientation is the same as on Fig.1a, namely agrees with experiments only at
velocities exceeding the critical one (of order of $cm/sec$ for
superconducting type II \textquotedblright low $T_{c}$ metals). Below it the
orientation is rotated by $30^{0}$.

The paper is organized as follows. Model is described, symmetries analyzed
and perturbative mean field solution developed in section II. The general
formalism is developed to treat the non Hermitian part of the equation. The
shape and the orientation of the vortex lattice and the reorientation
transition are described in section III. Then in section IV we calculate
corrections due to higher Landau levels and derive general expression for
conductivity. It is compared with a recent experiment. Section V is a
summary.

\section{Model and its perturbative flux flow solution}

\subsection{Time dependent GL Model}

Our starting point is the TDGL equation \cite{Tinkham}
\begin{equation}
\frac{{\hbar }^{2}\gamma }{2m_{ab}}\left( \frac{\partial }{\partial t}+\frac{%
ie^{\ast }}{\hbar }\Phi \right) \psi =-\frac{\delta }{\delta \psi ^{\ast }}F.
\label{fullTDGL}
\end{equation}%
The static GL free energy is:
\begin{equation}
F=\int d^{3}x\left( \frac{{\hbar }^{2}}{2m_{ab}}|(\vec{\nabla}+\frac{%
ie^{\ast }}{\hbar c}\vec{A})\psi |^{2}+\frac{{\hbar }^{2}}{2m_{c}}|\partial
_{z}\psi |^{2}-\alpha (T_{c}-T)|\psi |^{2}+\frac{b^{\prime }}{2}|\psi
|^{4}\right) ,  \label{energy}
\end{equation}%
where $\alpha $ and $b^{\prime }$ are phenomenological parameters, $\gamma $
is the inverse diffusion constant which controls the scale of dynamical
processes via dissipation. As usual the magnetic induction is $%
\overrightarrow{B}=\vec{\nabla}\times \vec{A}$ and electric field $%
\overrightarrow{E}=-\vec{\nabla}\Phi -\frac{\partial }{\partial t}\vec{A}.$
It should be supplemented by Amphere's law \cite{Hu,Troy}%
\begin{equation}
\vec{\nabla}\times \overrightarrow{B}=\sigma _{n}\overrightarrow{E}+%
\overrightarrow{J}_{s},
\end{equation}%
where the first term is the contribution of the normal liquid in the
framework of the two liquid model and the second term is the supercurrent
\begin{equation}
\overrightarrow{J}_{s}=-\frac{i{\hbar e}^{\ast }}{2m}\psi ^{\ast }\left(
\vec{\nabla}+\frac{ie^{\ast }}{\hbar c}\vec{A}\right) \psi +c.c.
\label{suppercurrent}
\end{equation}%
Tensor $\sigma _{n}$ is the normal state conductivity. We assume that the
coefficient of the covariant time derivative term $\gamma $ in eq.(\ref%
{fullTDGL}) is real although a small imaginary (Hall) part is always present
\cite{Troy}. The general case will be discussed in section V.

We make several approximations (identical to those made in \cite{Blum} and
major parts of \cite{Hu}) so that the problem becomes manageable. The
physical conditions allowing those approximations are the following.
Temperatures and magnetic fields are close \textquotedblleft
enough\textquotedblright\ to $H_{c2}(T)$. Under this assumption the order
parameter $\psi $ is suppressed compared to its Meissner value. In this
paper we will also assume strongly type II superconductivity $\kappa
=\lambda /\xi >>1$ ($\xi ^{2}={\hbar }^{2}/\left( 2m_{ab}\alpha T_{c}\right)
,$ $\lambda ^{2}=\frac{c^{2}m^{\ast }b^{\prime }}{4\pi e^{\ast 2}\alpha T_{c}%
}$). Magnetic field is very homogeneous since the vortices overlap.
Characteristic length describing the inhomogeneity of the electric field was
identified in \cite{Hu}: $\zeta ^{2}=\frac{4\pi \sigma _{n}}{\gamma }\lambda
^{2}$ and since typically\ $\sigma _{n}\simeq \gamma $, thus $\zeta >>\xi $
and the electric field is assumed homogeneous. Therefore the Maxwell type
equations for electromagnetic field are not considered. The time independent
vector potential will be taken in Landau gauge $\vec{A}=(By,0,0)$ and
describes a nonfluctuating magnetic field in the direction $-\widehat{z}$.
The scalar potential is also independent of time $A_{0}=Ey$ and describes
the electric field oriented along negative $y$ axis. The vortices are
therefore moving along the $x$ direction. We neglect thermal fluctuations
and disorder on the mesoscopic scale.

Throughout most of the paper we will use the following physical units. Unit
of length is the coherence length $\xi $, unit of magnetic field is $H_{c2}=%
\frac{\Phi _{0}}{2\pi \xi ^{2}}$, $\lambda =\frac{c}{e^{\ast }}\sqrt{\frac{%
m_{ab}b^{,}}{4\pi \alpha T_{c}}}$, and the unit of energy (temperature) is $%
T_{c}$. In these units the magnetic field is denoted by $b\equiv B/H_{c2}$.
The asymmetry of masses between the $z$ direction and the $x-y$ plane can be
removed by rescaling coordinates and time: $x\rightarrow \xi x/\sqrt{b}%
,y\rightarrow \xi y/\sqrt{b},z\rightarrow \xi z/\sqrt{bm_{c}/m_{ab}}%
,t\rightarrow \frac{\gamma \xi ^{2}}{2b}t$. The TDGL equations, after the
order parameter field is rescaled as well $\psi \rightarrow \sqrt{\frac{%
2\alpha T_{c}b}{b^{\prime }}}\psi ,$ is:%
\begin{eqnarray}
0 &=&L\psi +\psi |\psi |^{2},  \label{TDGLscaled} \\
L &\equiv &D_{t}-\frac{1}{2}\left[ D_{x}^{2}+\partial _{y}^{2}+\partial
_{z}^{2}\right] -a,  \notag
\end{eqnarray}%
where $a\equiv \frac{1-T/T_{c}}{2b},$ $v=\frac{c\gamma E}{2B}\sqrt{\frac{%
\hbar c}{e^{\ast }B}}$is scaled vortex velocity (in units of $2\sqrt{2\pi
B/\Phi _{0}}/\gamma $) and covariant derivatives are defined by $D_{x}=\frac{%
\partial }{\partial x}-iy$ and $D_{t}=\frac{\partial }{\partial t}+ivy$.
Since $\partial _{z}^{2}$ commutes with $L$, the equations are invariant
under the $z$ translations, the $z$ dependence of the solutions decouples
and is generally a plane wave. It is easy to see that the relevant solution
does not break this symmetry and is therefore constant with respect to $z$.
Consequently we consider the problem as a 2+1 dimensional one (note however
that if the 3D disorder or thermal fluctuations are included one can not
ignore the $z$ coordinate as the configuration of disorder can destroy the
translational symmetry along the $z$ direction) .

\subsection{Expansion of a nontrivial solution around dynamical phase
transition point}

The line\ in parameter space ($a,v$), which separates the normal region in
which the only solution is $\psi =0$ from the flux flow nontrivial solution
region, has been found by Hu and Thompson \cite{Hu}. We will construct a
perturbative solution of the TDGL equations near the mixed state - normal
phase transition line analogous to the one in statics \cite{Lascher}. The
range of applicability and precision of the LLL approximation at large $%
\kappa $ in statics was explored recently \cite{LiHLL}. The main difficulty
in the dynamical case is that the linear part of the equation $L$ is not
Hermitian due to the dissipation term $D_{t}$.

General idea of the expansion around a bifurcation point of a nonlinear
equation is as follows. One looks for a set of eigenfunctions of the linear
part of eq.( \ref{TDGLscaled}):
\begin{equation}
L_{Np\omega }\phi =\Theta _{Np\omega }\phi _{Np\omega }.
\end{equation}%
The operator $L$ consists of two parts: the usual Hermitian Hamiltonian of
particle in magnetic field $-\frac{1}{2}\left[ D_{x}^{2}+\partial _{y}^{2}%
\right] $ and the anti - hermitian covariant time derivative $D_{t}$. The
complete set of eigenfunctions with \textquotedblleft
quantum\textquotedblright\ numbers $N$ and $p_{x}\equiv p$ is:%
\begin{eqnarray}
\phi _{Np\omega } &=&\frac{1}{\sqrt{\pi 2^{N}N!}}\exp [i(px-\omega
t)]H_{N}(y-p+iv)\exp \left[ -\frac{1}{2}\left( y-p+iv\right) ^{2}\right]
\notag \\
\Theta _{Np\omega } &=&-a+N+\frac{1}{2}+\frac{v^{2}}{2}-i\left( \omega
-vp\right) ,  \label{HLL}
\end{eqnarray}%
where $H_{N}$ are Hermit polynomials. Unlike the usual case of a Hermitian
operator, eigenfunctions and eigenvalues of the Hermitian conjugate of the
operator $L^{\dag }$ are different:%
\begin{eqnarray}
L^{\dag }\phi _{Np\omega } &=&\Theta _{Np\omega }\phi _{Np\omega }  \notag \\
\overline{\phi }_{Np\omega } &=&\frac{1}{\sqrt{\pi 2^{N}N!}}\exp
[-i(px-\omega t)]H_{N}(y-p+iv)\exp \left[ -\frac{1}{2}\left( y-p+iv\right)
^{2}\right]  \notag \\
\overline{\Theta }_{Np\omega } &=&-a+N+\frac{1}{2}+\frac{v^{2}}{2}+i\left(
\omega -vp\right) .  \label{phibar}
\end{eqnarray}%
Note that $\overline{\phi }$ is \textit{not }a complex conjugate of $\phi $.
The orthogonality relations in the dynamical case involve both $\phi
_{Np\omega }$ and $\overline{\phi }_{Np\omega }$:%
\begin{eqnarray}
\int_{x,y,t}\overline{\phi }_{Np\omega }(x,y,t)\phi _{N^{\prime }p^{\prime
}\omega ^{\prime }}(x,y,t) &=&\left( 2\pi \right) ^{2}\delta _{NN^{\prime
}}\delta (p-p^{\prime })\delta (\omega -\omega ^{\prime }).  \label{ortho} \\
\left\langle \overline{\phi }_{Np\omega }(x,y,t)\phi _{Np\omega
}(x,y,t)\right\rangle _{x,y,t} &=&1,  \notag
\end{eqnarray}%
where the averaging over space and time is denoted by $\left\langle
...\right\rangle _{x,y,t}$.

The bifurcation (in this case the dynamical transition) occurs when there
exists a set of eigenfunctions of $L^{\prime }$ with zero eigenvalues $%
\Theta _{Np\omega }=0$:
\begin{eqnarray}
a_{bif}(v) &=&N+\frac{1}{2}+\frac{v^{2}}{2},  \label{phasetransition} \\
\omega &=&vp.  \label{manifold}
\end{eqnarray}%
It is clear that solutions with $N>0$ are unstable as in the static case
\cite{Tinkham}.\ Equation (\ref{phasetransition}) with $N=0$ gives the phase
transition line of \cite{Hu}, while eq.(\ref{manifold}) selects the
\textquotedblleft zero manifold\textquotedblright\ in the space of
functions. We define the \textquotedblleft distance\textquotedblright\ from
the transition line
\begin{equation}
a_{h}(v)\equiv a-a_{bif}(v)=a-\frac{1}{2}-\frac{v^{2}}{2}.  \label{ah}
\end{equation}

\bigskip When $a_{h}(v)>0$, the nonlinear TDGL equation%
\begin{eqnarray}
L\psi +\psi |\psi |^{2} &\equiv &L_{sh}\psi -a_{h}(v)\psi +\psi |\psi |^{2}=0
\label{x3half} \\
L_{sh} &=&L+a_{h}(v)  \notag
\end{eqnarray}%
is solved perturbatively in $a_{h}$ with a nonanalytic prefactor, as in the
static case:%
\begin{equation}
\Phi =\left[ a_{h}{}(v)\right] ^{1/2}\left[ \Phi ^{0}+a_{h}\Phi ^{1}+...%
\right] .  \label{fi}
\end{equation}%
To order $\left[ a_{h}{}\right] ^{1/2}$, the equation linearizes:
\begin{equation}
L_{sh}\Phi ^{0}=0.  \label{leading}
\end{equation}%
Therefore $\Phi _{0}$ belongs to the "zero manifold" and thereby can be
expanded :
\begin{equation}
\Phi ^{0}=\sum_{p}c_{p}\phi _{N=0,p,\omega =vp}\equiv \sum_{p}c_{p}\phi _{p},
\label{Phi}
\end{equation}%
with coefficients $c_{p}$ determined by the next order equation. As a
result, since all the $\phi _{p}(x,y,t)$ depend only on the combination $%
px-\omega t=p(x-vt)$ rather than separately on $x$ and $t,$ vortices \ move
in the direction perpendicular to both electric and magnetic field with
constant velocity $v.$ To order $\left[ a_{h}{}\right] ^{3/2}$, one obtains%
\begin{equation*}
L_{sh}\Phi ^{1}=\Phi ^{0}-\Phi ^{0}|\Phi ^{0}|^{2}.
\end{equation*}%
Multiplying this equation by $\overline{\phi }_{p}$ and integrating over
\bigskip $(x,y,t)$ using of the orthogonality relation eq.(\ref{ortho}) one
obtains the following infinite set of nonlinear algebraic equations:%
\begin{equation}
\sum_{p_{1},p_{2},r}c_{p_{1}}c_{p_{2}}c_{r}^{\ast }\left\langle \overline{%
\phi }_{p}\phi _{r}^{\ast }\phi _{p_{1}}\phi _{p_{2}}\right\rangle
_{x,y,t}=c_{p}.  \label{cLeqs}
\end{equation}%
We will study the solution of this set in the next section.

\section{ Shape and orientation of the moving lattice.}

\subsection{Symmetry and energetics considerations}

It is well known in the static case that there is a solution of GL equations
for any lattice symmetry. The same is true in the dynamical case as well,
but the symmetries should take into account the motion of vortices. Define
the covariant derivatives in a matrix 2+1 dimensional form (summation over
repeated indices assumed):%
\begin{equation}
A_{\mu }=b_{\mu \nu }x_{\nu };\text{ \ \ }D_{\mu }=\partial _{\mu }-iA_{\mu
}.
\end{equation}%
and the Landau gauge
\begin{equation}
b_{\mu \nu }=%
\begin{tabular}{|l|l|l|}
\hline
$0$ & $1$ & $0$ \\ \hline
$0$ & $0$ & $0$ \\ \hline
$0$ & $-v$ & $0$ \\ \hline
\end{tabular}%
\end{equation}%
is used in our paper. All indices run over space $\mu =1$ $(x)$, $2(y)$ and $%
3(t)$. The electromagnetic translation operators satisfying $[T_{d},D_{\mu
}]=0$ are:%
\begin{equation}
T_{d}=e^{i\mathbf{d\cdot P}}=\exp \left[ -i\left( \frac{1}{2}d_{\mu }b_{\mu
\nu }d_{\nu }+x_{\nu }b_{\nu \mu }d_{\mu }\right) \right] e^{i\mathbf{d\cdot
p}},
\end{equation}%
where generators are $P_{\mu }=-i\left( \partial _{\mu }-ib_{\nu \mu }x_{\nu
}\right) $ (note transpose in the matrix $b_{\mu \nu }$). Operators $p_{\mu
}=-i\partial _{\mu }$ are usual (not "electromagnetic") translation
operators. The following commutation relations
\begin{equation}
\left[ P_{\mu },P_{\nu }\right] =i\left( b_{\mu \nu }-b_{\nu \mu }\right) .
\end{equation}%
can be verified. Thus we will have $[i\mathbf{d}_{1}\mathbf{\cdot P},i%
\mathbf{d}_{2}\mathbf{\cdot P}]=-id_{1\alpha }d_{2\beta }\left( b_{\alpha
\beta }-b_{\beta \alpha }\right) $. Using the Haussdorf formula one checks
that the electromagnetic translation operators obey $e^{i\mathbf{d}_{1}%
\mathbf{\cdot P}}e^{i\mathbf{d}_{2}\mathbf{\cdot P}}=e^{i\mathbf{d}_{2}%
\mathbf{\cdot P}}e^{i\mathbf{d}_{1}\mathbf{\cdot P}}e^{\left[ i\mathbf{d}_{1}%
\mathbf{\cdot P},\ \ i\mathbf{d}_{2}\mathbf{\cdot P}\right] }$. If $\mathbf{d%
}_{1}$ and $\mathbf{d}_{2}$ are the lattice vectors which preserve the
symmetry of the system (when one translates system by $\mathbf{d}_{1}$ or $%
\mathbf{d}_{2}$, the system will be unchanged), one shall require $e^{i%
\mathbf{d}_{1}\mathbf{\cdot P}}\psi =e^{i\mathbf{d}_{2}\mathbf{\cdot P}}\psi
$ and it will lead to
\begin{equation*}
e^{i\mathbf{d}_{1}\mathbf{\cdot P}}e^{i\mathbf{d}_{2}\mathbf{\cdot P}}\psi
=e^{i\mathbf{d}_{2}\mathbf{\cdot P}}e^{i\mathbf{d}_{1}\mathbf{\cdot P}}e^{%
\left[ i\mathbf{d}_{1}\mathbf{\cdot P},\ \ i\mathbf{d}_{2}\mathbf{\cdot P}%
\right] }\psi =e^{\left[ i\mathbf{d}_{1}\mathbf{\cdot P},\ \ i\mathbf{d}_{2}%
\mathbf{\cdot P}\right] }\psi =\psi .
\end{equation*}%
Therefore we should demand%
\begin{equation*}
\left[ i\mathbf{d}_{1}\mathbf{\cdot P},\ \ i\mathbf{d}_{2}\mathbf{\cdot P}%
\right] =i2\pi \times \func{integer}.
\end{equation*}%
This requirement is satisfied by the following basic translation symmetry
vectors%
\begin{eqnarray}
\mathbf{d}^{(1)} &=&a_{\Delta }(\frac{1}{2},0,-\frac{1}{2v})  \notag \\
\mathbf{d}^{(2)} &=&a_{\Delta }(\frac{r}{2},r^{\prime },-\frac{r}{2v})
\label{symmetries} \\
\mathbf{d}^{(0)} &=&\tau (v,0,1).  \notag
\end{eqnarray}%
Here $a_{\Delta }$ is the lattice spacing along the direction of motion, $%
\tau $ is arbitrary (a continuous translational symmetry). The flux
quantization (one flux quantum per unit cell assumed) determines $r^{\prime
} $: $r^{\prime }a_{\Delta }^{2}=2\pi .$ The $\mathbf{d}^{(1)}$ translation
symmetry leads to discrete parameter%
\begin{equation*}
p=\frac{2\pi }{a_{\Delta }}l\equiv gl
\end{equation*}%
in eq.(\ref{Phi}), and the set of equations eq.(\ref{cLeqs}) will take a form%
\begin{eqnarray}
c_{n} &=&\sqrt{\frac{1}{2}}g^{2}%
\sum_{l_{1},l_{2}}c_{l_{1}+n}c_{l_{2}+n}c_{l_{1}+l_{2}+n}^{\ast }\times \\
&&\exp \left\{ -\frac{1}{2}\left[ \left( gl_{1}+iv\right) ^{2}+\left(
gl_{2}+iv\right) ^{2}-v^{2}\right] \right\} .  \notag
\end{eqnarray}%
It can be solved as in the static case by an Ansatz:
\begin{equation*}
c_{l}=\sqrt{\frac{g}{\sqrt{\pi }\beta _{A}(v)}}e^{-i\pi rl(l+1)}
\end{equation*}%
with the Abrikosov function
\begin{eqnarray}
\beta _{A}(v) &=&\frac{g}{\sqrt{2\pi }}\sum_{l1,l2}\exp \left\{ 2\pi
irl_{1}l_{2}\right\} \times  \label{betaA} \\
&&\exp \left\{ -\frac{1}{2}\left[ \left( gl_{1}+iv\right) ^{2}+\left(
gl_{2}+iv\right) ^{2}-v^{2}\right] \right\} .  \notag
\end{eqnarray}%
Consequently%
\begin{equation}
\Phi ^{_{0}}(x,y,z)=\frac{1}{\sqrt{\beta _{A}(v)}}\varphi (x,y),
\end{equation}%
where%
\begin{equation}
\varphi (x,y)\equiv \sqrt{\frac{g}{\sqrt{\pi }}}\sum_{l}\exp \left[
il(g(x-vt)-\pi r(l+1))\right] \exp \left[ -\frac{1}{2}\left( y-gl-iv\right)
^{2}\right]
\end{equation}%
is normalized by $\left\langle \left\vert \varphi \right\vert
^{2}\right\rangle _{x,y}=1$.

In the static case a solution which has minimal free energy is physically
realized. The free energy is proportional to $-\left[ a_{h}{}(0)\right]
^{2}/(2\beta _{A}\left( 0\right) )$ which therefore should be minimized.
This means that one should minimize $\beta _{A}\left( 0\right) .$ The
minimal $\beta _{A}\left( 0\right) =1.16$ is obtained for the hexagonal
lattice. Similar reasoning cannot be applied to the moving lattice solution
of the TDGL equation since the friction force is non conservative. Under
these circumstances Ketterson and Song \cite{Ketterson} calculated the work
made by the friction force:%
\begin{equation}
\overset{\cdot }{S}\equiv \frac{d}{dt}S=2\gamma \left\langle \left\vert
D_{t}\psi \right\vert ^{2}\right\rangle _{x,y}.
\end{equation}%
The preferred lattice structure in the steady state corresponds to a state
with largest $\overset{\cdot }{S}$. For the lattice solution of TDGL
equation one obtains to leading order in $\alpha _{h}$:%
\begin{eqnarray}
\overset{\cdot }{S} &\propto &\frac{g\left\vert \alpha _{h}(v)\right\vert }{%
\beta _{A}(v)}\times  \notag \\
&&\left\langle \left\vert \sum_{l}\left( \frac{\partial }{\partial t}%
+ivy\right) \exp \left[ il(g(x-vt)-\pi r(l+1))\right] \exp \left[ -\frac{1}{2%
}\left( y-gl-iv\right) ^{2}\right] \right\vert ^{2}\right\rangle _{x,y} \\
&=&\frac{v^{2}\left\vert \alpha _{h}(v)\right\vert }{2\beta _{A}(v)}%
e^{v^{2}}.  \notag
\end{eqnarray}%
We therefore shall minimize $\beta _{A}$ as function of $r$ and $a_{\Delta }$%
. This is consistent with the static case.

\subsection{The stationary orientation of the flux lattice. The
reorientation transition at high flux flow velocity\protect\bigskip}

We found that the minimum of $\beta _{A}(v)$ appears always when $r=1/2$,
namely for rhombic lattices. Therefore from now on we consider these
lattices only. As a function of an angle of the rhombic lattice $\tan \theta
=4\pi /a_{\Delta }^{2}$ (see Fig. 1 for definition of $\theta $) it
generally has two minima, see Fig. 3. In the static case the two minima are
degenerate with $\theta =60^{\circ }$, $30^{\circ }$ corresponding to
perpendicular orientations of the hexagonal lattice, while for nonzero
velocity the degeneracy is lifted. Note that originally \cite{Maki,Hu}\ it
was assumed that the lattice is hexagonal also in the dynamical case.
Generally the shape is not strongly distorted for physically realizable
velocities. For velocities smaller than $v_{c}=0.95$ angle $\theta $ close
to $60^{\circ }$ (the orientation\ of Fig.1b) is preferred over the one
close to $30^{\circ }$ (the orientation of Fig.1a), see Fig.3c. The
dependence of the angle $\theta $ on velocity can be very well fitted in the
whole range $v<0.5$ by%
\begin{equation}
\theta =30-0.4v-24v^{2}.
\end{equation}%
The Abrikosov function also depends on velocity increasing according to%
\begin{equation}
\beta _{A}(v)=\beta _{A}(0)(1+1.25v^{2}),  \label{beta}
\end{equation}%
where $\beta _{A}(0)=1.1596$ is the static value for hexagonal lattice. As
the critical velocity is approached the two minima coincide, see Fig.3b.
Beyond that point the preferred structure is just the opposite, Fig.3a. The
transition is first order and the coexistence region should exist.

We now make A few comments about the orientation of the lattice. The reader
might have noticed that the orientation of the lattice is not completely
arbitrary since direction of the vector $\mathbf{d}_{1}$ coincides with the
direction of the vortex motion. The most general $\beta _{A}(v)$ is given by
eq.(\ref{betaA}) with arbitrary $r$. We minimized numerically the Abrikosov $%
\beta $ function and found that the solution with the largest dissipation is
always of the more symmetric type $r=1/2$. One can argue that despite the
fact that electric field breaks the continuous rotational symmetry, it still
preserves a discrete transformation $y\rightarrow -y,\psi \rightarrow \psi
^{\ast }$. The solution $r=1/2$ preserves this discrete symmetry. This
symmetry is unlikely to be spontaneously broken. Indeed the symmetry was
observed in the experiments (for example, in \cite{Pardo}).

\section{Nonlinear conductivity and breakdown of the LLL scaling in transport%
\protect\bigskip}

In this section we first calculate the leading higher Landau level
corrections to the solution of the TDGL equation eq.(\ref{TDGLscaled}). Then
we use it to derive the correction to the LLL scaling of conductivity \cite%
{Ikeda,Troy,Blum}.

\subsection{Higher orders in $a_{h}$ correction to the moving lattice
solution}

Using the same symmetry arguments as for the leading order, the second term
in eq.(\ref{fi}) can be expanded as:
\begin{eqnarray}
\Phi ^{1} &=&\sum_{N=0}C_{N}^{1}\varphi _{N}  \notag \\
\varphi _{N} &=&\sqrt{\frac{g}{\sqrt{\pi }2^{N}N!}}\sum_{l}\exp \left[
il(g(x-vt)-\pi r(l+1))\right] \\
&&\times H_{N}(y-gl-iv)\exp \left[ -\frac{1}{2}\left( y-Gl-iv\right) ^{2}%
\right] .  \notag
\end{eqnarray}%
Multiplying eq.(\ref{x3half}) by $\overline{\varphi }_{N}$ for $N>0$, one
obtains:%
\begin{equation}
NC_{N}^{1}=-\beta _{A}^{-3/2}\left\langle \overline{\varphi }_{N}\varphi
^{\ast }\varphi \varphi \right\rangle
\end{equation}%
To find $C_{0}^{1}$ we need in addition also the order $a_{h}^{5/2}$
equation:
\begin{equation}
L_{sh}\Phi ^{2}=\Phi ^{1}-(2\Phi ^{1}|\Phi ^{0}|^{2}+\Phi ^{1\ast }\Phi
^{0}\Phi ^{0})  \label{eq2}
\end{equation}%
Inner product with $\varphi $ gives:
\begin{equation}
NC_{0}^{1}=-\beta _{A}^{-5/2}\sum_{N=1}^{\infty }\left[ 2\left\langle
\overline{\varphi }_{N}\varphi ^{\ast }\varphi \varphi \right\rangle
\left\langle \overline{\varphi }\varphi ^{\ast }\varphi _{N}\varphi
\right\rangle +\left\langle \overline{\varphi }_{N}^{\ast }\varphi ^{\ast
}\varphi ^{\ast }\varphi \right\rangle \left\langle \overline{\varphi }%
\varphi _{N}^{\ast }\varphi \varphi \right\rangle \right] .  \label{g1}
\end{equation}%
Note that for hexagonal lattice $\left\langle \overline{\varphi }_{N}\varphi
^{\ast }\varphi \varphi \right\rangle \neq 0$ only when $N=6j$, where $j$ is
an integer. This is due to hexagonal symmetry of the vortex lattice \cite%
{Lascher}. In statics $\beta _{N}=\left\langle \overline{\varphi }%
_{N}\varphi ^{\ast }\varphi \varphi \right\rangle $ decreases very fast with
$j$: $\beta _{6}=-.2787,\beta _{12}=.0249$ \cite{LiHLL}. Because of this the
coefficient of the next to leading order is very small (also an additional
factor of $6$ in the denominator helps the convergency).

\subsection{\protect\bigskip The LLL scaling in nonlinear conductivity}

In the flux flow regime, in addition to the normal state conductivity, there
is a large contribution from the Cooper pairs represented by the order
parameter field. It was noted in \cite{Ikeda,Troy,Blum} that the LLL
contribution to nonlinear conductivity

\begin{equation}
\sigma =-\frac{i{\hbar e}^{\ast }}{2mE}\psi ^{\ast }\left( \vec{\nabla}+%
\frac{ie^{\ast }}{\hbar c}\vec{A}\right) \psi
\end{equation}%
is proportional to the superfluid density. The scaled dimensionless
conductivity is defined as $\sigma _{scaled}\equiv \frac{4\pi \kappa ^{2}}{%
c^{2}\gamma }\sigma $ and $\sigma _{scaled}$ in LLL approximation is:
\begin{equation}
\sigma _{LLL}=\frac{i}{2v}\left\langle \Psi _{LLL}^{\ast }\partial _{y}\Psi
_{LLL}-\Psi _{LLL}\partial _{y}\Psi _{LLL}^{\ast }\right\rangle
=\left\langle \left\vert \Psi _{LLL}\right\vert ^{2}\right\rangle .
\end{equation}%
The last equality is due to the general property of the LLL functions, see
eq.(\ref{HLL}). It follows naive expectation of "Drude" like formula \cite%
{Tinkham} with $\left\vert \Psi _{LLL}\right\vert ^{2}$ playing a role of
"charge carriers" density (meaning here Cooper pairs).

To leading order in $a_{h}$ using the results of the section II one gets%
\begin{equation}
\sigma _{LLL}=\frac{ia_{h}(v)}{2\beta _{A}(v)v}\left\langle \varphi ^{\ast
}\partial _{y}\varphi -\varphi \partial _{y}\varphi ^{\ast }\right\rangle =%
\frac{a_{h}(v)}{\beta _{A}(v)}e^{v^{2}},
\end{equation}%
where $a_{h}(v)=(1-t_{GL}-b-v^{2})/(2b)$. At finite $v$ there is an
exponential factor coming from the nonorthogonality of eigenfunctions of a
non Hermitian operator and, in addition, similar dependence in $\beta _{A}$
and quadratic in $a_{h}$. In the limit $v\rightarrow 0$ one recovers the
Ohmic expression (see \cite{Troy}) returning to standard units
\begin{equation}
\sigma _{LLL}^{(1)}=\sigma _{0}\frac{1-t_{GL}-b}{2b\beta _{A}(0)},\text{ \ \
}\sigma _{0}\equiv \frac{c^{2}\gamma }{4\pi \kappa ^{2}},  \label{linear}
\end{equation}%
while the leading nonlinear (cubic) correction is, using eq.(\ref{beta}),%
\begin{equation}
\sigma _{LLL}^{(3)}=\text{\ }\sigma _{0}\frac{t_{GL}+b-5}{8b\beta _{A}(0)}%
v^{2}.  \label{nonlinear}
\end{equation}%
\ where $v=\frac{c\gamma E}{2B}\sqrt{\frac{\hbar c}{e^{\ast }B}}$.

\subsection{ Leading correction to the LLL scaling}

Generally to all orders in $a_{h}$ one can write $\Psi
=\sum_{N}C_{N}(a_{h})\varphi _{N}$
\begin{equation}
\sigma =\text{\ }\sigma _{0}\frac{i}{2v}\sum_{NM}C_{N}^{\ast
}(a_{h})C_{M}(a_{h})\left\langle \varphi _{N}^{\ast }\partial _{y}\varphi
_{M}-\varphi _{M}\partial _{y}\varphi _{N}^{\ast }\right\rangle \equiv \text{%
\ }\sigma _{0}\sum_{NM}C_{N}^{\ast }(a_{h})C_{M}(a_{h})\sigma _{NM}
\end{equation}%
For $N>M$ and $N-M$ even integer%
\begin{equation}
\sigma _{NM}=-\text{\ }\sqrt{\frac{2^{N-M}M!}{N!}}(-v^{2})^{\frac{N-M}{2}%
}[v^{2}L_{M-1}^{N-M+1}(-2v^{2})+\frac{M+1}{2}%
L_{M+1}^{N-M-1}(-2v^{2})]e^{v^{2}},
\end{equation}%
where $L(y)$ are Laguerre polynomials. This contribution is always subohmic $%
\sigma _{NM}\sim v^{N-M}$ at small $v$. If $N-M$ is odd, the contribution
vanishes. The diagonal contributions are simpler
\begin{equation}
\sigma _{NN}=\text{\ }[L_{N-1}^{1}(-2v^{2})+L_{N}^{1}(-2v^{2})]e^{v^{2}}
\end{equation}%
and have an ohmic part
\begin{equation*}
\sigma _{NN}=2N+1.
\end{equation*}%
The first term, proportional to Landau orbital number $N,$ is responsible
for the breaking of the naive "Drude" like expectation that conductivity is
proportional to $\left\vert \Psi \right\vert ^{2}$ \cite{Tinkham}$.$ One
observes that higher Landau levels contribute to conductivity more than to $%
\left\vert \Psi \right\vert ^{2}$. One can interpret this as "increased
charge movers density".

Thus the ohmic conductivity has two contributions:%
\begin{eqnarray}
\sigma ^{(1)} &=&\text{\ }\sigma _{0}\sum_{N}(2N+1)\left\vert
C_{N}(a_{h})\right\vert ^{2}=\sigma _{1}+\sigma _{2}; \\
\sigma _{1} &=&\sigma _{0}\left\langle \left\vert \Psi \right\vert
^{2}\right\rangle ;\text{ \ \ \ }\sigma _{2}=2\text{\ }\sigma
_{0}\sum_{N=1}N\left\vert C_{N}(a_{h})\right\vert ^{2},
\end{eqnarray}%
the first proportional to the superfluid density, while the second, the HLL
part, is not and is of order $a_{h}^{3}$ only. Substituting expressions for $%
C_{0}$ from the previous section, we obtain for the Ohmic conductivity to
order $a_{h}^{2}$
\begin{equation}
\sigma _{1}=\text{\ }\sigma _{0}\left[ \frac{a_{h}}{\beta _{A}}+3\frac{%
a_{h}^{2}}{\beta _{A}^{3}}\sum_{N=1}\frac{\beta _{N}^{2}}{N}\right] ;
\label{condHLL}
\end{equation}%
where all the quantities are taken in the limit $v\rightarrow 0$. The sum
rapidly converges in the static (or low velocity) case: $\sum_{N}\frac{\beta
_{N}^{2}}{N}=0.0131$.

\subsection{Comparison with experiment}

On Fig. 4 we compare the results with a recent experiments at high currents
(electric fields) of \cite{Sulpice} on $Nb$ in which vortex velocities as
high as $10^{5}$ $cm/sec$. We used the same values of the Ginzburg - Landau
parameter $\kappa =9.4$ and the inverse diffusion constant $\gamma =1.17$ $%
sec/cm^{2}$ to fit all three curves corresponding to magnetic fields $H=80$ $%
mT,$ $100$ $mT$ and $120$ $mT$ for "cold" sample with $T_{c}=8.6$ $K$. We
used the measured (inset on Fig.2 of \cite{Sulpice}) $H_{c2}\equiv T_{c}%
\frac{dH_{c2}(T)}{dT}|_{T=T_{c}}=4.4$ $T$. The temperature was $T=7.8$ $K$
close enough to $T_{c}$ so that the $a_{h}^{2}$ correction was always below $%
10\dot{\%}$. The value of parameter $\gamma $ is in good agreement
the measured normal state resistivity of$\ 9.9$ $\mu \Omega $
$cm$. The results agree well with the flux flow Ohmic conductivity
data at relatively low currents (still well above the critical
current) exhibiting the $1/H$ behavior presented in Fig.2 of
\cite{Sulpice}.

One observes that the full expression (solid lines) is closer to the
experiment at very high electric fields. Several curves for magnetic field
... are given. The smallest is clearly off the LLL approach range.

\section{Conclusion}

To summarize, we have considered the dynamics of vortex lattice neglecting
effects of pinning. We studied the time dependent Ginzburg - Landau equation
in the lowest Landau level approximation. The validity region of the LLL
approximation, as in the static case, we require $a_{h}=\frac{1}{2B}\left[
H_{c2}(T)-B-\frac{c^{2}\gamma ^{2}\Phi _{0}E^{2}}{4\pi B^{2}}\right] <<6,$
the factor $6$ coming from cancellations of the higher Landau level effects
due to hexagonal symmetry (even the hexagonal symmetry is approximate in the
moving lattice). We systematically calculated higher Landau level
corrections to conductivity and the vortex lattice energy dissipation. The
stationary lattice structure depends on the flux flow velocity. While for
small velocities $V<2v_{c}\sqrt{2\pi B/\Phi _{0}}/\gamma ,v_{c}=0.95$ vortex
lattice is oriented as in Fig.1b, while beyond this velocity orients like in
Fig.1a. We emphasize that in our calculation the pinning effect was
disregarded. Of course, as was firmly established in numerous theoretical
and experimental investigations, pinning significantly can modify the
picture for low velocities. Pinning generally \textquotedblleft
prefers\textquotedblright\ configuration of Fig.1a and this is a possible
reason why the experimental observed orientation is depicted as Fig.1a.
However one can expect that for higher velocities and very clean samples the
pinning dominated dynamics crosses over into the interactions dominated
dynamics considered in the present work. The high velocity of order $cm/sec$
is unlikely to be seen in decoration experiments. However other techniques
like SANS and muon spin rotation \cite{Forgan} and possibly Lorentz
microscopy \cite{Tonomura} are able to detect the lattice structure even at
such relatively high velocities. At very high velocities the results for
nonlinear conductivity agree with recent experiments \cite{Sulpice}.

\bigskip

\acknowledgments We are especially grateful to F.P. Lin for fitting the
conductivity data. We are grateful to E. Andrei, A. Knigavko, T.J. Yang, A.
Shaulov, Y. Yeshurun for discussion and E.M. Forgan for discussion and
sending results prior to publication . The work of A.M. was supported by the
graduate students program of NCTS. Work of B.R. is supported by NSC of
R.O.C., NSC\#932112M009024 and hospitality of Peking University. Work of
D.L. is supported by the Ministry of Science and Technology of China
(\#G1999064602) and national natural science foundation of China
(\#10274030).

\begin{center}
{\Huge Figure captions}
\end{center}

{\LARGE Fig. 1}

Two possible orientations of the (approximately) hexagonal vortex lattice.
(a) direction of the flux lines is the same as the nearest neighbors lattice
orientation. (b) direction of the flux lines is perpendicular to the nearest
neighbors lattice orientation.

{\LARGE Fig.2}

The flux lattice geometry: $\mathbf{d}^{(1)},\mathbf{d}^{(2)}$ are the
translational \ symmetry vectors which determines the primitive cell of the
flux lattice. The angle between these two vectors is $\theta .$

{\LARGE Fig.3 }

Dependence of the Abrikosov $\beta $ parameter on orientation and shape of
the vortex lattice moving with scaled velocities $v=0.5.0.95,1.1$. The angle
$\theta $ is defined as an angle between the direction of motion and a
crystallographic axis in direction of the symmetry transformation $d_{2}$.
The minimum favors the smaller angle close to $30^{0}$ corresponding to
structure of the Fig.1a for $v<v_{c}$, while the other local minimum
corresponding to Fig1b (angles close to $60^{0}$) is preferred for $v>v_{c}$.

{\LARGE Fig.4 }

Current - voltage curves at high flux flow velocities. Data of ref. \cite%
{Sulpice} on $Nb$ films at $T=7.8$ $K$ (symbols represent different magnetic
fields) are compared with theory combining the linear (Ohmic) contribution
eq.(\ref{condHLL}) and the cubic correction eq.(\ref{nonlinear}).

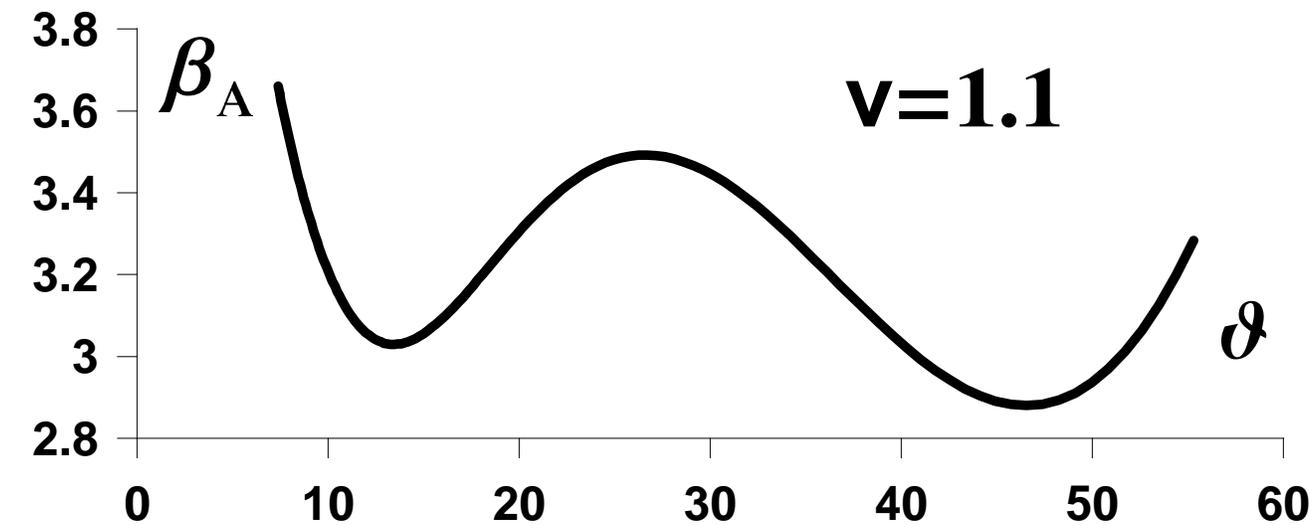
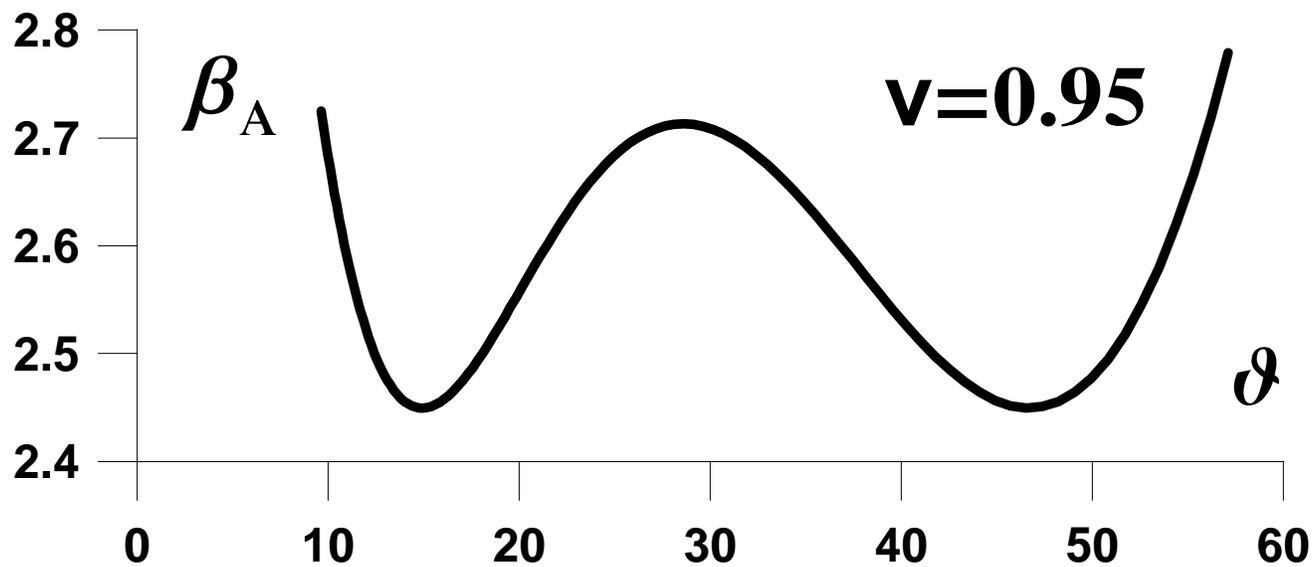
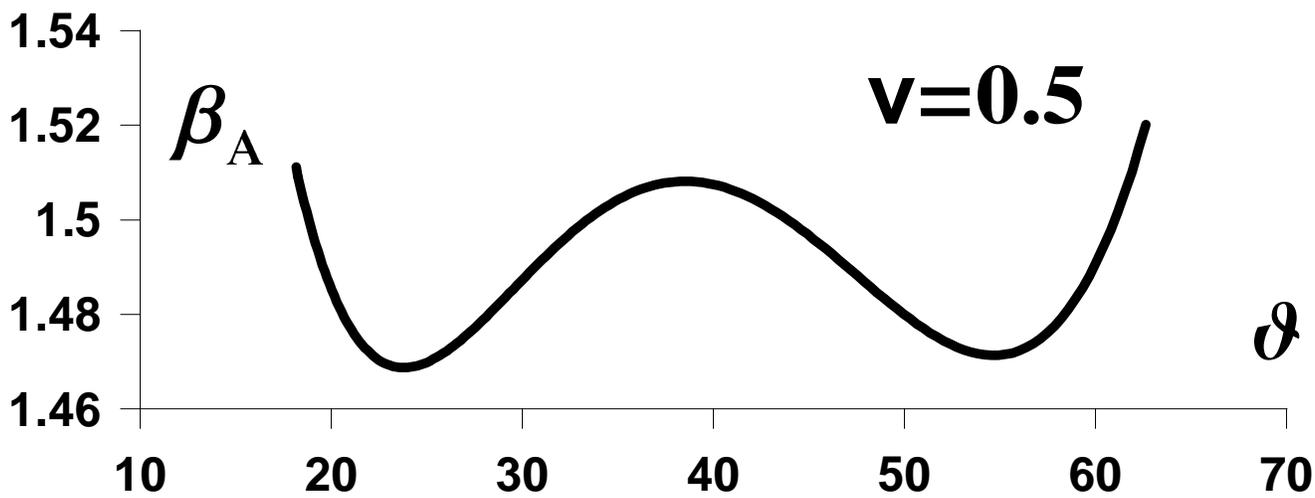
\end{document}